\documentclass[superscriptaddress,
amsmath,amssymb,
aps,prc,reprint,
]{revtex4-2}

\usepackage{multirow,graphicx,url,braket}
\usepackage{amssymb}
\usepackage{amsmath}
\usepackage{dcolumn}
\usepackage{bm}
\usepackage{hyperref}
\usepackage[ulem=normalem]{changes}
\usepackage{threeparttable}
\begin{document}

\title{Quantifying neutron-proton interactions in $N=51$ isotones: from NEEC candidate $^{93}$Mo to $^{99}$Cd}

\author{B.~Maheshwari}
\email{bhoomika.physics@gmail.com}
\affiliation{Grand Accélérateur National d'Ions Lourds,
CEA/DSM-CNRS/IN2P3, Bvd Henri Becquerel, BP 55027, F-14076 Caen, France}

\author{P.~Van~Isacker}
\affiliation{Grand Accélérateur National d'Ions Lourds,
CEA/DSM-CNRS/IN2P3, Bvd Henri Becquerel, BP 55027, F-14076 Caen, France}

\author{P.~M.~Walker}
\affiliation{Department of Physics, University of Surrey, Guildford, GU2 7XH, United Kingdom}

\date{\today}

\begin{abstract}

We present a shell-model analysis of $N=51$ isotones, $^{93}$Mo, $^{95}$Ru, $^{97}$Pd, and $^{99}$Cd, to quantify the role of neutron-proton interactions in shaping the location and half-life of isomeric states. The study is motivated by the anomalous behavior of the ${21/2}^+$ isomeric state in $^{93}$Mo, a prominent candidate for nuclear excitation by electron capture (NEEC), which misses an $E2$ decay branch due to a higher-lying ${17/2}^+$ state and instead proceeds via a long-lived $E4$ isomeric transition. Employing a consistent configuration space and empirically derived effective interaction, we extract and compare the proton-proton and neutron-proton matrix elements for the four $N=51$ isotones. Our results show a distinct dominance of the neutron-proton interaction in $^{93}$Mo, in contrast to its neighbors--$^{95}$Ru, $^{97}$Pd, and $^{99}$Cd--where no analogous isomeric behavior emerges due to structural evolution. These findings reveal that the favorable structure for NEEC in $^{93}$Mo stems from subtle interaction systematics that do not persist across the chain. We find that the $E2$ strength of the key NEEC transition is reduced by 40\% compared to the previously estimated value. The analysis provides microscopic insights into the origin of long-lived isomerism in medium-mass nuclei and outlines a framework for identifying future candidates in other mass regions for exploiting the potential energy storage capacities of isomeric states.

\end{abstract}

\maketitle


\section{Introduction} 
\label{sec:intro}

Nuclear isomers remain a subject of topical discussion due to their interdisciplinary applications in fundamental  physics, nucleosynthesis, nuclear medicine, and potential energy-storage capacities~\cite{isomer}. Among these applications, induced isomeric depletion, a process that holds significant potential for controlled energy storage, has been demonstrated in at least eight cases via different processes operating at the intersection of nuclear and atomic physics~\cite{carroll2024}. One of such process, nuclear excitation by electron capture (NEEC), was originally proposed in 1976~\cite{goldanski1976}, but its application to isomeric states was first suggested in 2002~\cite{zadernovksy2002}. Only recently, the first evidence for the NEEC activity of the 6.85 hour, ${21/2}^+$ isomeric state in $^{93}$Mo $(Z=42, N=51)$ was reported~\cite{chiara2018}
though different experimental conditions may lead to different results~\cite{guo2022}. Besides the difficulties in observing the NEEC mechanism, a principal concern is the theoretical interpretation of the NEEC probability~\cite{wu2019} which is orders of magnitude smaller than observed. 

\begin{figure}[!htb]
\centering
\resizebox{0.48\textwidth}{!}{\includegraphics[width=\textwidth]{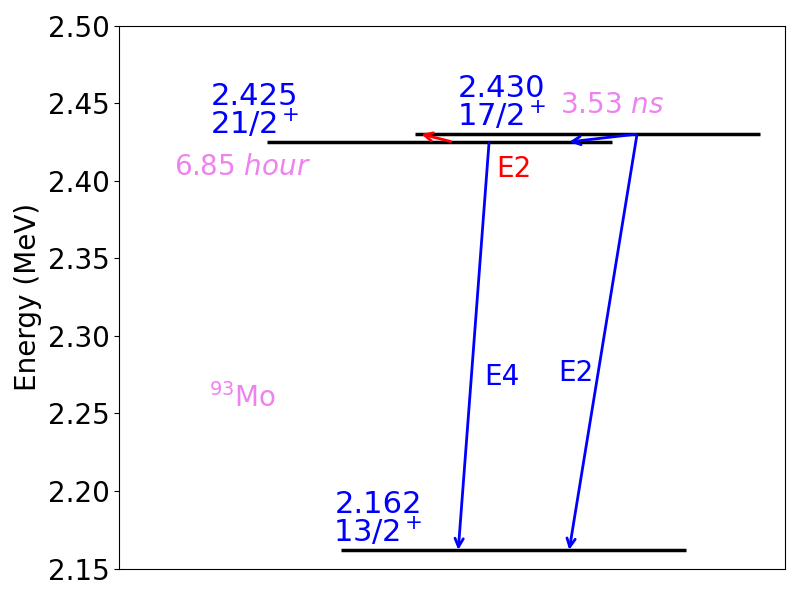}}
\caption{(Color online)
Long-lived ${21/2}^+,E4$ isomeric decay due to the higher-lying ${17/2}^+$ state in $^{93}$Mo. Induced isomeric depopulation $({21/2}^+ \rightarrow {17/2}^+)$ is shown in red.}
\label{fig:93mois}\end{figure}

The theoretical estimation of the probability of NEEC directly depends on the probabilities of nuclear transitions from the involved ${21/2}^+$ isomeric state which lies below the ${15/2}^+$, ${17/2}^+$, and ${19/2}^+$ states for $^{93}$Mo, leading to the possibility of induced isomeric decay as shown in Fig.~\ref{fig:93mois}. One of the key transitions decisive for the observation of NEEC in $^{93}$Mo is a 5-keV $ {17/2}^+ \rightarrow {21/2}^+$ transition for which the $B(E2)$ value is not measured yet. So far, only a theoretical $B(E2)$ estimate is known for this transition which is based on large-scale shell-model calculations using the pairing-plus-quadrupole interaction with monopole corrections designed to reproduce the energy levels of nuclides with $40 \le Z \le 42$ and $50 \le N \le 53$~\cite{hasegawa2011}. 

The unusual inversion of the ${21/2}^+$ and ${17/2}^+$ states in $^{93}$Mo, contrary to the expected seniority $v=3$, $\pi g_{9/2}^3$ structure in $^{93}$Tc, was first explained by Auerbach and Talmi in 1964~\cite{auerbach1964} using the $\nu 1d_{5/2} \otimes \pi 0g_{9/2}^2  $ configuration, though no $E2$ transition-rate estimates were provided. The experimental $g$-factor for the ${21/2}^+$ isomeric state also suggested a fully aligned $\nu 1d_{5/2} \otimes \pi 0g_{9/2}^2  $  configuration~\cite{kaindl1973}. Subsequent large-scale shell model calculations~\cite{hasegawa2011} again attributed this inversion to strong neutron-proton ($\nu\pi$) interactions, though the complexity of two-body matrix elements in such high-dimensional configurations complicates the quantitative analysis. 

The ${21/2}^+$ isomeric state also exists in the next neighboring $N=51$ isotone, $^{95}$Ru, with a much shorter half-life of 10.05 nanoseconds, since the ${17/2}^+$ state lies below the isomeric state enabling a faster $E2$ transition. Its primary configuration was also found to be $\nu 1d_{5/2} \otimes \pi 0g_{9/2}^2  $  by Ghugre \textit{et al.}~\cite{ghugre1994} using Gloeckner's interaction~\cite{gloeckner1975} for this region considering a $^{88}$Sr $(Z=38, N=50)$ core but no $E2$ transition-rate estimates were provided. Galindo \textit{et al.}~\cite{galindo2004} also confirmed the same for $^{95}$Ru until the ${29/2}^+$ state with no need of particle-hole excitations which were otherwise found to be essential for the higher spins, and they also listed $E2$ transition probabilities. Hasegawa \textit{et al.}~\cite{hasegawa2011} also validated the level ordering of the ${17/2}^+$ and ${21/2}^+$ states in $^{95}$Ru, though no $E2$ transition probabilities were quoted.   

No ${21/2}^+$ isomers are yet reported in the heavier $N=51$ isotones, $^{97}$Pd, and $^{99}$Cd. This work examines the evolution of $\nu\pi$ matrix elements in the $N=51$ isotones based on the dominant $\nu 1d_{5/2} \otimes \pi 0g_{9/2}^n $ configuration, where $n$ is the number of valence protons with a $Z=40$ closed shell configuration. No $E2$ transition-rate estimates for these $N=51$ isotones, using a primary $\nu 1d_{5/2} \otimes \pi 0g_{9/2}^n  $ configuration, have been reported in the literature to our knowledge, but they are essential for a quantitative understanding of the role of the $\nu\pi$ interaction. Further theoretical as well as experimental studies are necessary for an understanding of the evolution of nuclear structure as the proton number increases from $Z=42$ to $Z=48$. This analysis will also help the search for isomers in other regions of the nuclear chart, which are due to spin inversion as a result of the $\nu\pi$ interaction and which may be potential NEEC candidates.      

\section{Formalism}
\label{sec:formalism}
We perform shell-model calculations for the $N=51$ isotones
$^{93}$Mo, $^{95}$Ru, $^{97}$Pd, and $^{99}$Cd.
Based on the spin-parity of the ground states of these nuclei,
the odd neutron most likely sits in the $1d_{5/2}$ orbital.
Therefore, a dominant configuration of $\nu1d_{5/2}\otimes\pi0g_{9/2}^n$
can be assumed for the low-lying states, including the ${21/2}^+$ isomer.
We study in particular the energy difference $E_{\rm x}({21/2}^+)-E_{\rm x}({17/2}^+)$
as the valence proton number ($n$, with respect to the $Z=40$ closed shell) increases.
The shell-model Hamiltonian can in general be written as
\begin{equation}
\hat H=\sum_i\epsilon_i\hat n_i+\frac{1}{2}\sum_{i\neq j}\langle ij|\hat V|ij\rangle+\cdots,
\label{eq:ham}
\end{equation}
where the first term refers to single-particle energies
and the second term denotes the two-body interaction.
The model space $\nu1d_{5/2}\otimes\pi0g_{9/2}^n$ is sufficiently simple
for obtaining closed analytic expressions of the matrix elements between many-body states
in terms of the single-particle energies $\epsilon_i$ and the two-body interaction.
The interaction $\hat V_{\rho\rho}$ between like particles
($\rho=\nu$ for neutrons and $\rho=\pi$ for protons,
the latter applying to the case considered here)
can be calculated with two distinct methods~\cite{talmi}.
The first relies on a recursive procedure
that expresses a matrix element between $n$-particle states
in terms of those between $(n-1)$-particle states,
\begin{widetext}
\begin{equation}
\langle j^n\alpha J|\hat V_{\rho\rho}|j^n\alpha'J\rangle=
\frac{n}{n-2}\sum_{\alpha_1\alpha'_1J_1}
[j^{n-1}(\alpha_1J_1)jJ|\}j^n\alpha J]
[j^{n-1}(\alpha'_1J_1)jJ|\}j^n\alpha'J]
\langle j^{n-1}\alpha_1J_1|\hat V_{\rho\rho}|j^{n-1}\alpha'_1J_1\rangle,
\label{eq:recur1}
\end{equation}
where $[j^{n-1}(\alpha_1J_1)jJ|\}j^n\alpha J]$ is an $n\rightarrow n-1$ coefficient of fractional parentage (CFP).
There can be more than one state of $n$ particles in a single-$j$ orbital with total angular momentum $J$,
in which case an additional quantum number is needed, denoted here as $\alpha$.
Alternatively, with the second method,
one calculates the $n$-body matrix element directly from the expression
\begin{equation}
\langle j^n\alpha J|\hat V_{\rho\rho}|j^n\alpha'J\rangle=
\frac{n(n-1)}{2}\sum_{\alpha_2J_2J'}
[j^{n-2}(\alpha_2J_2)j^2(J')J|\}j^n\alpha J]
[j^{n-2}(\alpha_2J_2)j^2(J')J|\}j^n\alpha'J]
V_{\rho\rho}^{J'},
\label{eq:recur2}
\end{equation}
in terms of the $n\rightarrow n-2$ CFPs $[j^{n-2}(\alpha_2J_2)j^2(J')J|\}j^n\alpha J]$
and the two-body matrix elements $V_{\rho\rho}^{J'}\equiv\langle j^2J'|\hat V_{\rho\rho}|j^2J'\rangle$.
In the analysis that follows the neutron-proton ($\nu\pi$) interaction plays a crucial role.
For the present case with one neutron in $1d_{5/2}$ and $n$ protons in $0g_{9/2}$
the following expression applies for the $\nu\pi$ interaction matrix element:
\begin{align}
&\langle j_\nu j_\pi^n(\alpha J_\pi);J|\hat V_{\nu\pi}|j_\nu j_\pi^n(\alpha'J'_\pi);J\rangle
\nonumber\\&=
n[J_\pi][J'_\pi]\sum_{\alpha_1J_1J'}(2J'+1)
[j_\pi^{n-1}(\alpha_1J_1)j_\pi J_\pi|\}j_\pi^n\alpha J_\pi]
[j_\pi^{n-1}(\alpha_1J_1)j_\pi J'_\pi|\}j_\pi^n\alpha'J'_\pi]
\Bigl\{\begin{array}{ccc}
j_\pi&j_\nu&J'\\J&J_1&J_\pi
\end{array}\Bigr\}
\Bigl\{\begin{array}{ccc}
j_\pi&j_\nu&J'\\J&J_1&J'_\pi
\end{array}\Bigr\}
V_{\nu\pi}^{J'},
\label{eq:npmat}
\end{align}
\end{widetext}
where the notation $[J]\equiv\sqrt{2J+1}$ is used.
In the present case, with a neutron in $1d_{5/2}$ and protons in $0g_{9/2}$,
the angular momentum $J'$ in the $\nu\pi$ two-body interaction runs from 2 to 7.
The two-body matrix elements for the interaction among the protons in $0g_{9/2}$
can be empirically derived, as discussed in Ref.~\cite{ley2023}.
The energy spectra of $^{92}$Mo and $^{98}$Cd are used, 
corresponding to two valence particles and two valence holes in $0g_{9/2}$, respectively,
and an interpolated interaction is taken for intermediate $N=50$ isotones.
Such empirically derived interaction works well for $N=50$ isotones~\cite{ley2023}
and explains the origin of seniority isomers, though particle-hole symmetry is broken.
The $\pi\pi$ interaction matrix elements are summarised in the first two rows of Table~\ref{tab:int}.
The $\nu1d_{5/2}$--$\pi0g_{9/2}$ $\nu\pi$ interaction is fixed from the energy spectrum of $^{92}$Nb.
The interaction $V_{\nu\pi}^J$ is strongest for the fully aligned state with $J=7$,
which is the ground state of $^{92}$Nb.
The last row in Table~\ref{tab:int} lists the matrix elements
between a neutron particle in $1d_{5/2}$ and a proton hole in $0g_{9/2}$ utilizing Pandya's transformation~\cite{pandya1956},
and in this case the interaction is strongest for $J=6$.
Note that for each set of matrix elements we have put the most attractive one to zero,
which can be done if one is only interested in excitation energies 
and not in absolute energies. The same holds for the single-particle energies for our current configuration with only one neutron and one proton orbital; they do not affect the calculated excitation energies. 
\begin{table}
\centering
\caption{\label{tab:int}
Interaction matrix elements (in MeV) for neutrons in $1d_{5/2}$ and protons in $0g_{9/2}$.}
\begin{ruledtabular}
\begin{tabular}{c|cccccc}
Interaction&$J$&0&2&4&6&8\\
\hline
$V_{\pi\pi}^J(0g_{9/2}0g_{9/2})$&$^{92}$Mo&0.000&1.510&2.283&2.612&2.761\\
$V_{\pi\pi}^J(0g_{9/2}0g_{9/2})$&$^{98}$Cd&0.000&1.395&2.083&2.281&2.428\\
\hline
&2&3&4&5&6&7\\
\hline
$V_{\nu\pi}^J(1d_{5/2}0g_{9/2})$&$0.135$&0.286&0.480&0.357&0.501&0.000\\
$V_{\nu\pi^{-1}}^J(1d_{5/2}0g_{9/2}^{-1})$&$0.757$&0.185&0.156&0.036&0.000&0.219\\
\end{tabular}
\end{ruledtabular}
\end{table}

The shell-model expressions remain relatively simple for $^{93}$Mo
with one valence neutron and two valence protons,
in which case the elements of the Hamiltonian matrix
can be obtained from the expressions given above.
Complexities increase as one includes more valence protons.
The elements of the Hamiltonian matrix can then be obtained symbolically as a function
of the single-particle energies and two-body matrix elements
with the Mathematica code {\tt shell.m}~\cite{isackercode}.
The eigenvalue problem, however, requires a numerical diagonalization.

Reduced electric transition probabilities
between an initial state $\alpha_{\rm i}J_{\rm i}$ and a final state $\alpha_{\rm f}J_{\rm f}$,
induced by the electric transition operator $\hat O(ELM)$,
can be expressed as
\begin{equation}
B(EL;\alpha_{\rm i}J_{\rm i}\rightarrow\alpha_{\rm f}J_{\rm f})=
\frac{|\langle\alpha_{\rm f}J_{\rm f}||\hat O(EL)||\alpha_{\rm i}J_{\rm i}\rangle|^2}{2J_{\rm i}+1},
\label{eq:bela}
\end{equation}
where $\langle\cdot||\cdot||\cdot\rangle$ is a reduced matrix element~\cite{talmi}
that does not depend on the projection quantum numbers of the angular momentum.
Equation~(\ref{eq:bela}) applies to many-body eigenstates $\alpha J$,
obtained after diagonalization of the Hamiltonian matrix,
which are expanded in a chosen shell-model basis such that 
\begin{align}
|\alpha_{\rm i}J_{\rm i}M_{\rm i}\rangle={}&
\sum_k a_k(\alpha_{\rm i}J_{\rm i})|kJ_{\rm i}M_{\rm i}\rangle,
\nonumber\\
|\alpha_{\rm f}J_{\rm f}M_{\rm f}\rangle={}&
\sum_l b_l(\alpha_{\rm f}J_{\rm f})|lJ_{\rm f}M_{\rm f}\rangle,
\label{eq:expan}
\end{align}
where $k$ ($l$) labels basis states with angular momentum $J_{\rm i}$ ($J_{\rm f}$).
Combination of Eqs.~(\ref{eq:bela}) and~(\ref{eq:expan})
leads to a generalized expression for the reduced transition probability,
\begin{align}
&B(EL;\alpha_{\rm i}J_{\rm i}\rightarrow\alpha_{\rm f}J_{\rm f})
\nonumber\\&=
\frac{1}{2J_{\rm i}+1}
\biggl|\sum_{kl}a_k(\alpha_{\rm i}J_{\rm i})b_l(\alpha_{\rm f}J_{\rm f})\langle lJ_{\rm f}
||\hat O(EL)||kJ_{\rm i}\rangle\biggr|^2,
\label{eq:belb}
\end{align}
This can explain the interference effects between neutrons and protons,
leading to smaller or larger reduced transition probabilities
as compared to those associated with the basis states.

In a $\nu\pi$ basis the matrix elements of a one-body operator with a neutron and a proton part
can be calculated with the general expression
\begin{widetext}
\begin{align}
\langle\alpha_\nu J_\nu,\alpha_\pi J_\pi;J||\hat O(EL)||\alpha'_\nu J'_\nu,\alpha'_\pi J'_\pi;J'\rangle={}&
(-)^{J_\nu+L}[J][J']\Bigl[
(-)^{J_\pi+J'}
\Bigl\{\begin{array}{ccc}
J_\nu&J&J_\pi\\J'&J'_\nu&L
\end{array}\Bigr\}
\langle\alpha_\nu J_\nu||\hat O_\nu(EL)||\alpha'_\nu J'_\nu\rangle
\delta_{\alpha_\pi\alpha'_\pi}\delta_{J_\pi J'_\pi}
\nonumber\\&+
(-)^{J'_\pi+J}
\Bigl\{\begin{array}{ccc}
J_\pi&J&J_\nu\\J'&J'_\pi&L
\end{array}\Bigr\}
\langle\alpha_\pi J_\pi||\hat O_\pi(EL)||\alpha'_\pi J'_\pi\rangle
\delta_{\alpha_\nu\alpha'_\nu}\delta_{J_\nu J'_\nu}\Bigr].
\label{eq:elm}
\end{align}
In the application to the $N=51$ isotones
there is only one neutron in the $1d_{5/2}$ valence orbital
and the reduced matrix element of the neutron part of the $EL$ operator
is a single-particle reduced matrix element, which is discussed below.
The reduced matrix element of the proton part of the $EL$ operator
is between $n$-proton states
and can be calculated with the formula
\begin{align}
&\langle j^n\alpha J||\hat O(EL)||j^n\alpha'J'\rangle
\nonumber\\&=
n[J][J']
\sum_{\alpha_1J_1}
(-)^{J_1+j+J+L}
[j^{n-1}(\alpha_1J_1)jJ|\}j^n\alpha J]
[j^{n-1}(\alpha_1J_1)jJ'|\}j^n\alpha'J']
\Bigl\{\begin{array}{ccc}
j&J&J_1\\J'&j&L
\end{array}\Bigr\}
\langle{\cal N}lj||\hat O(EL)||{\cal N}lj\rangle,
\end{align}
\end{widetext}
where the last term denotes the single-particle reduced matrix element
of the operator $\hat O(EL)=e\sum_ir_i^LY_L(\theta_i,\phi_i)$,
with $e$ the effective charge of the nucleon.
Note that this matrix element not only depends on the single-particle angular momentum $j$
but also on the principal quantum number $\cal N$ and the orbital angular momentum $l$.
Standard expressions are available for the single-particle reduced matrix elements~\cite{talmi},
which for a neutron in $1d_{5/2}$ and for a proton in $0g_{9/2}$ lead to
\begin{align}
\langle\nu1d_{5/2}||\hat O_\nu(E2)||\nu1d_{5/2}\rangle={}&-11\sqrt{\frac{3}{7\pi}}e_\nu b^2,
\nonumber\\
\langle\pi0g_{9/2}||\hat O_\pi(E2)||\pi0g_{9/2}\rangle={}&-5\sqrt{\frac{11}{3\pi}}e_\pi b^2,
\label{eq:e2sp}
\end{align}
where $e_\nu$ and $e_\pi$ are the neutron and proton effective charges.
The length parameter of the harmonic oscillator is $b$,
for which we take $b^2=41.46/(45A^{-1/3}-25A^{-2/3})$~${\rm fm}^2$~\cite{blomqvist1968},
resulting in $B(E2)$ values in units of $e^2{\rm fm}^4$.
The proton effective charge $e_\pi$ is set to 1.32~\cite{ley2023},
and three sets of neutron effective charge $e_\nu$ are adopted:
i) 1.18 from the quadrupole moment $Q(5/2^+)$ of the $5/2^+$ ground state of $^{91}$Zr~\cite{nndc},
ii) 1.73 from the $B(E2;2_1^+\rightarrow0_1^+)$ value in $^{92}$Zr~\cite{nndc},
and iii) the averaged value 1.48.
Further, all $B(E2)$ values can be expressed in single-particle units, also called Weisskopf units (W.u.),
\begin{equation}
B_{\rm W}(E2)=0.0594\times A^{4/3}e^2{\rm fm}^4, 
\label{eq:weiss}
\end{equation}
where $A$ is the mass number. 

For comparison we will also show results of a large-scale shell-model (LSSM) calculation
for the $N=51$ isotones under consideration.
The GLEKPN interaction~\cite{mach1990} is used
in the valence space $20\leq Z\leq50$ and $40\leq N\leq70$.
The active neutron orbitals are
$0g_{9/2}$, $0g_{7/2}$, $1d_{5/2}$, $1d_{3/2}$, and $2s_{1/2}$
with respective single-particle energies of $-10.164$, $+11.927$, $+4.220$, $+7.017$, and $+4.371$ MeV
while the proton orbitals include
$0f_{7/2}$, $0f_{5/2}$, $1p_{3/2}$, $1p_{1/2}$, and $0g_{9/2}$
with respective single-particle energies of $-8.085$, $-10.840$, $-11.361$, $-9.610$, $-4.424$ MeV.
The neutron and proton effective charges 
adopted in the LSSM calculation are $e_\nu=1.178$ and $e_\pi=1.315$.
The calculations are performed with a $^{78}$Ni core
for reasons of computational feasibility,
and in this space the shell-model Hamiltonian
is diagonalized with use of the block Lanczos method of KSHELL~\cite{shimizu2019}.
The harmonic oscillator parameter is chosen to be $\hbar\omega=45A^{-1/3}-25A^{-2/3}$~MeV
and $\beta_{\rm cm}=100$~MeV to eliminate spurious states due to the center-of-mass motion.

\section{Discussion}
\label{sec:discussion}

\subsection{The nucleus $^{93}$Mo}
\label{subs:mo93}
\begin{figure*}
\centering
\includegraphics[width=0.89\textwidth]{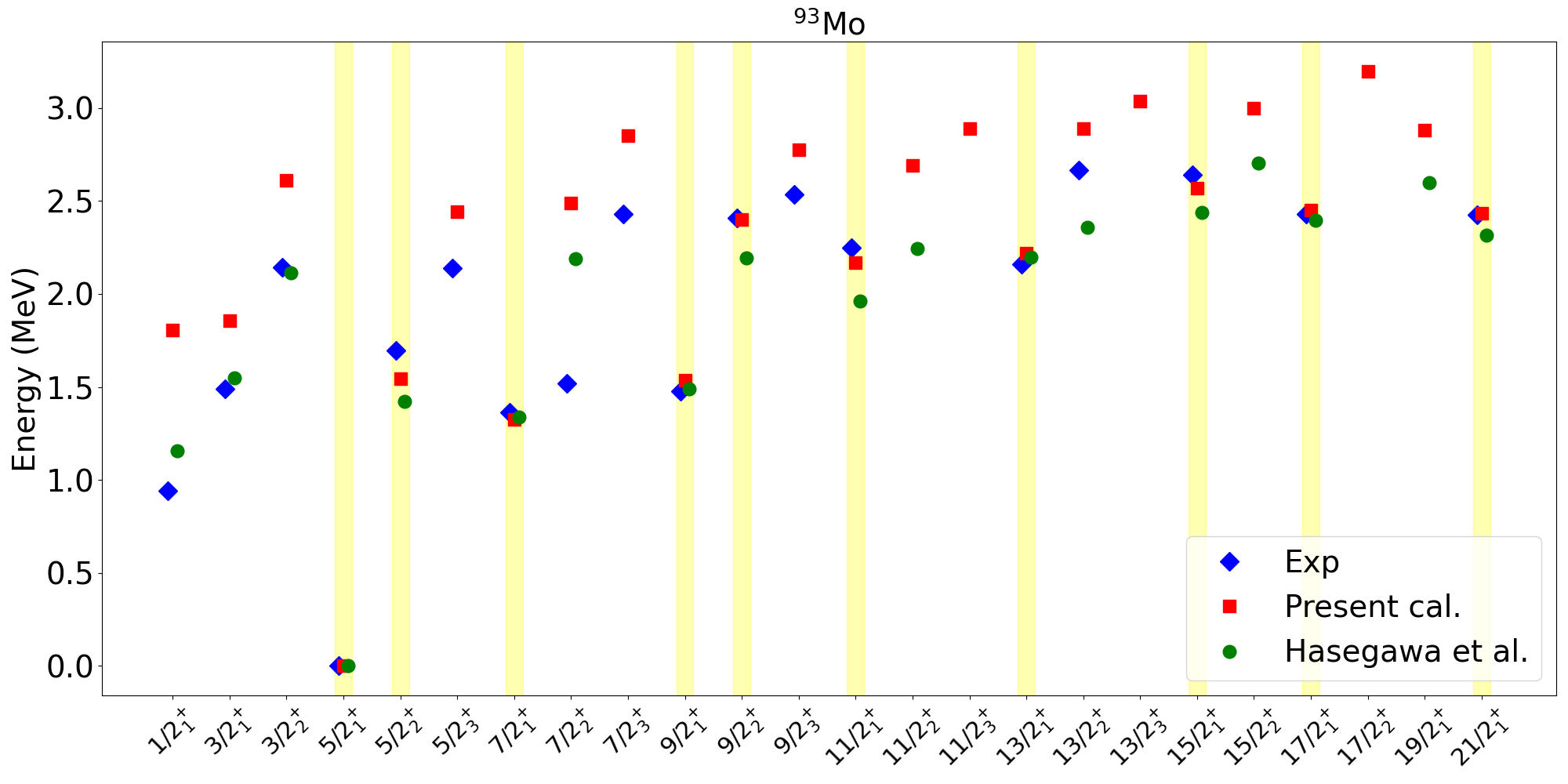}
\caption{\label{fig:enmo93}(Color online)
Experimental~\cite{nndc} and calculated energies of various lower-lying states in $^{93}$Mo.
Theoretical results are obtained in the $\nu1d_{5/2}\otimes\pi0g_{9/2}^2$ model space (present)
or taken from Hasegawa {\it et al.}~\cite{hasegawa2011}. 
The yellow shades highlight very good agreement ($\pm 0.2$~MeV) of the present calculation with the data.}
\end{figure*}

\begin{table}[!htb]
\caption{\label{tab:enmo93}
Experimental~\cite{nndc} and calculated energies (in MeV) in $^{93}$Mo.
Theoretical results are obtained in the $\nu1d_{5/2}\otimes\pi0g_{9/2}^2$ model space (present),
taken from Hasegawa {\it et al.}~\cite{hasegawa2011}, or calculated with the GLEKPN interaction.}
\centering
\begin{tabular}{|ccccc|}
\hline
$J^\pi$ & Exp & Present & Ref.~\cite{hasegawa2011}& GLEKPN  \\
\hline
${1/2}^+$ & 0.943 & 1.807 & 1.155 & 1.073 \\
${3/2}^+$ & 1.492 & 1.859 & 1.549 & 1.192 \\
    & 2.145 & 2.612 & 2.113 & 2.388\\
${5/2}^+$ & 0 & 0 & 0 & 0 \\
    & 1.695 & 1.546 & 1.422 & 1.912 \\
    & 2.141 & 2.445 & --- & 2.311\\
${7/2}^+$ & 1.363 & 1.326 & 1.339& 1.309\\
    & 1.520 & 2.490 & 2.188 & 1.503\\
    & 2.430 & 2.853 & --- & 2.580\\
${9/2}^+$ & 1.477 & 1.539 & 1.492 & 1.810 \\
    & 2.409 & 2.401 & 2.192 & 2.581\\
    & 2.534 & 2.775 & --- & 2.807\\
${11/2}^+$ & 2.247 & 2.171 & 1.960 & 2.449\\
    & --- & 2.693 & 2.245 & 2.805 \\
    & --- & 2.889 & --- & 2.871\\
${13/2}^+$ & 2.161 & 2.221 & 2.197 & 2.453\\
    & 2.667 & 2.893 & 2.359 & 3.186\\
    & --- & 3.036 &--- & 3.402\\
${15/2}^+$ & 2.641 & 2.571 & 2.438 & 2.807 \\
    & --- & 3.000 & 2.704 & 3.441 \\
${17/2}^+$ & 2.429 & 2.454 & 2.398 & 2.766 \\
    & --- & 3.197 & --- & 3.696 \\
${19/2}^+$ & --- & 2.880 & 2.600 & 3.371\\
${21/2}^+$ & 2.424 & 2.437 & 2.315 & 2.788\\
\hline
\end{tabular}
\end{table}
In $^{93}$Mo, with a single neutron in $\nu1d_{5/2}$ and two protons in $\pi0g_{9/2}$,
the allowed states are $1/2^+(1)$, $3/2^+(2)$, $5/2^+(3)$, $7/2^+(3)$, $9/2^+(3)$, $11/2^+(3)$,
$13/2^+(3)$, $15/2^+(2)$, $17/2^+(2)$, $19/2^+(1)$, $21/2^+(1)$,
with the number in brackets referring to multiplicity of the state.
This case is simple enough to investigate the quantitative role of the $\nu\pi$ interaction
leading to a $21/2^+$ isomer located below the $17/2^+$ state.

\begin{table*}[!htb]
\caption{\label{tab:emmo93}
Experimental~\cite{nndc} and calculated reduced $E2$ transition probabilities in $^{93}$Mo.
Expressions for the $B(E2)$ values are given in units of $e^2b^4$,
where $b$ is length parameter of the harmonic oscillator,
or in Weisskopf units (W.u.) when numerical values are used for the effective charges.
Also shown are the results of Hasegawa \textit{et al.}~\cite{hasegawa2011}
and of the LSSM calculation with the GLEKPN interaction.
The bold $B(E2)$ values correspond to the transitions important in the mechanism of isomeric depletion.}
\centering
\resizebox{0.75\textwidth}{!}{
\begin{tabular}{|c|c|c|c|c|c|c|c|}
\hline
&&Expression&
\multicolumn{3}{c|}{Present calculation}&Ref.~\cite{hasegawa2011}&GLEKPN\\
$J_{\rm i}\rightarrow J_{\rm f}$&Exp&(in units of $e^2b^4$)&
\multicolumn{3}{c|}{$e_\pi=1.32$}&$e_\pi=1.50$&$e_\pi=1.315$\\
\cline{4-6}
&&&$e_\nu=1.18$&$e_\nu=1.78$&$e_\nu=1.48$&
$e_\nu=0.50$&$e_\nu=1.178$\\
\hline
${17/2}^+ \rightarrow {21/2}^+$ & --- & $(0.23\,e_\nu+0.93\,e_\pi)^2$ &
$\textbf{2.02}$& $\textbf{2.40}$ & $\textbf{2.21}$ & $\textbf{3.5}$ & $\textbf{2.9}$  \\
${17/2}^+ \rightarrow {13/2}^+$ & \textbf{4.48(23)} & $(0.28\,e_\nu+1.28\,e_\pi)^2$ & \textbf{3.67} &\textbf{ 4.30} & \textbf{3.98} & \textbf{4.0} & \textbf{5.0} \\
${19/2}^+ \rightarrow {15/2}^+$ & --- & $(0.015\,e_\nu-0.83\,e_\pi)^2$&  1.04 & 1.03 & 1.03 & 2.5 & 0.4\\
${19/2}^+ \rightarrow {17/2}^+$ & --- & $(0.29\,e_\nu+0.019\,e_\pi)^2$ & 0.09 & 0.22 & 0.15 & 0.01 & 0.0\\
${19/2}^+ \rightarrow {21/2}^+$ & --- & $(1.09\,e_\nu-0.58\,e_\pi)^2$ & 0.25 & 1.26 & 0.66 & 0.03 & 0.7 \\
${15/2}^+ \rightarrow {17/2}^+$ & --- & $(1.02\,e_\nu-0.032\,e_\pi)^2$ & 1.23 & 2.87 & 1.96 & 0.02 & 1.7 \\
${15/2}^+ \rightarrow {11/2}^+$ & --- & $(0.22\,e_\nu+1.24\,e_\pi)^2$ & 3.23 & 3.69 & 3.46 & 2.2 & 2.0 \\
${15/2}^+ \rightarrow {13/2}^+$ & --- & $(0.37\,e_\nu+0.36\,e_\pi)^2$& 0.75 & 1.15 & 0.94 & 0.01 & 0.0 \\
${13/2}^+ \rightarrow {9/2}^+$ & 3.3(5) & $(0.20\,e_\nu+1.53\,e_\pi)^2$ & 4.59 & 5.09 & 4.84 & 7.2 & 3.6 \\
${9/2}^+ \rightarrow {7/2}^+$ & 90(80) & $(0.87\,e_\nu+1.25\,e_\pi)^2$& 6.50 & 9.28 & 7.83 & 6.5 & 1.4 \\
${9/2}^+ \rightarrow {5/2}^+$ & 12(4) & $(0.088\,e_\nu+1.38\,e_\pi)^2$& 3.35 & 3.53 & 3.44 & 12.9 & 3.9 \\
${7/2}^+ \rightarrow {5/2}^+$ & 8.7(25) & $(0.21\,e_\nu+1.68\,e_\pi)^2$& 5.49 & 6.06 & 5.77 & 12.6 & 3.5 \\
${3/2}^+ \rightarrow {1/2}^+$ & --- & $(0.93\,e_\nu-0.33\,e_\pi)^2$& 0.39 & 1.34 & 0.80 & 0.51 & 5.5 \\
${1/2}^+ \rightarrow {5/2}^+$ & $80^{+40}_{-80}$ & $(0.22\,e_\nu+1.25\,e_\pi)^2$ & 3.28 & 3.74 & 3.51 & 25 & 10.9 \\
${3/2}^+ \rightarrow {5/2}^+$ & --- & $(0.24\,e_\nu-1.30\,e_\pi)^2$ &  1.87 & 1.52 & 1.69 & 4.9 & 3.1\\
\hline
\end{tabular}}
\end{table*}

\begin{table*}[!htb]
\caption{\label{tab:enrupdcd}
Experimental~\cite{nndc} and calculated energies (in MeV) in $^{95}$Ru, $^{97}$Pd, and $^{99}$Cd.
Theoretical results are obtained in the $\nu1d_{5/2}\otimes\pi0g_{9/2}^n$ model space (present)
for $n=4$, 6, and 8, respectively, or calculated with the GLEKPN interaction.}
\centering
\resizebox{0.75\textwidth}{!}
{$\begin{tabular}{|c|ccc|ccc|ccc|}
\hline
& \multicolumn{3}{c|}{$^{95}$Ru} & \multicolumn{3}{c|}{$^{97}$Pd} & \multicolumn{3}{c|}{$^{99}$Cd} \\
\hline
$J^\pi$ & Exp & Present & GLEKPN & Exp & Present & GLEKPN & Exp & Present & GLEKPN \\
\hline
$1/2^+$ & 0.787 & 1.539 & 0.659 &  0.774 & 1.393 & 0.370 & 0.882 & 1.277 & 0.624\\
$3/2^+$ & --- & 1.666 & 0.863 & --- & 1.486 & 0.739 & 1.076 & 1.303 & 1.165 \\
$5/2^+$ & 0 & 0  & 0 & 0 & 0& 0.178 & 0 & 0& 1.005\\
    & --- & 1.605  & 1.806 & 1.712 & 1.620 & 1.619 & 1.781 & 1.610 & 1.454 \\
    & --- & 2.283 & 2.204 &  1.782 & 2.174& 1.686 & --- & 2.304& 2.025\\
$7/2^+$ & 0.941 & 1.383 & 0.579 & 0.686 & 1.443 &0 &  0.440 & 1.497 & 0\\
    & --- & 2.299 & 1.458 & 1.043 & 2.236 & 1.478 & 1.606 & 2.229 & 1.970\\
$9/2^+$ & 1.352  & 1.439 & 1.502 & 1.294 & 1.363 & 1.495 &  1.224 & 1.273 & 1.737\\
    & 2.117 & 2.144 & 2.125 & 2.176 & 2.139 & 1.750 & 1.974 & 2.125 & 2.157\\
    & 2.294 & 2.485 & 2.351 & 2.505 & 2.493 & 2.219 & 2.226 & 2.739 & 2.737\\
$11/2^+$ & 2.067? & 2.118 & 2.116 & 1.943 & 2.104 & 1.492 &  1.674 & 2.086 & 1.405\\
    & 2.246 & 2.246 & 2.237 & 2.395 & 2.289 & 2.068 & --- & 2.326 & 2.447\\
    & --- & 2.672 & 2.304& 2.890 & 2.525 & 2.423& --- & 3.132 & 2.585\\
$13/2^+$ & 2.029 & 2.132 & 2.197 & 1.881 & 1.995 & 2.109 &  1.831 & 1.924 & 2.402 \\
    & --- & 2.301 & 2.440 & 2.141 & 2.186 & 2.289 & --- & 2.322 & 2.730\\
    & --- & 2.716 & 2.958 & 2.481 & 2.379 & 2.399 & --- & 2.719 & 2.927\\
$15/2^+$ & --- & 2.431 & 2.492 & 2.371 & 2.360 & 2.212 & --- & 2.226 & 2.159\\
    & --- & 2.648 & 2.855 & 2.500 & 2.416 & 2.313 & --- & 2.392 & 2.736\\
$17/2^+$ & 2.284 & 2.329 & 2.446 & 2.244 & 2.280 & 2.469 &  2.057 & 2.165 & 2.557 \\
$19/2^+$ & 3.362 &  2.597 & 2.782 & 2.469 & 2.428 & 2.484 & 2.274 & 2.306 & 2.375\\
$21/2^+$ & 2.538 & 2.477 & 2.950 & 2.639 & 2.495 & 3.040 & 2.700 & 2.514 & 3.029 \\
    & --- & 3.939 & 3.712 & 3.578 & 3.616 & 3.313 & --- & --- & 4.342\\
$25/2^+$ & 3.830 & 3.899 & 4.534 & 3.810 & 3.828 & 4.553 & --- & --- & --- \\
$29/2^+$ & 4.502 & 4.690 & 5.342 & 4.636 & 4.648 & 5.759 & --- & --- & --- \\
\hline
\end{tabular}$}
\end{table*}

The $21/2^+$ state can only be generated with $J_\pi=8$ in $|\nu1d_{5/2}\otimes\pi0g_{9/2}^2(J_\pi);21/2\rangle$.
From Eq.~(\ref{eq:npmat}) the following $\nu\pi$ matrix element is obtained
\begin{equation}
\textstyle
V_{\nu \pi}(21/2^+)=\frac{5}{14}V^6_{\nu\pi}+\frac{23}{14}V^7_{\nu \pi},
\label{eq:mo21a}
\end{equation}
and therefore the $J=21/2^+$ energy is given by
\begin{equation}
\textstyle
E(21/2^+)=E_{\rm sp}+
V^8_{\pi\pi}+\frac{5}{14}V^6_{\nu \pi}+\frac{23}{14}V^7_{\nu \pi},
\label{eq:mo21b}
\end{equation}
where $E_{\rm sp}$ is a short-hand notation for the single-particle energy,
which for $^{93}$Mo equals $E_{\rm sp}=\epsilon_{1d_{5/2}}+2\epsilon_{0g_{9/2}}$.
There are two  $|\nu1d_{5/2}\otimes\pi0g_{9/2}^2(J_\pi);17/2\rangle$ states,
with $J_\pi=6$ and $J_\pi=8$, respectively,
leading to the following Hamiltonian matrix
\begin{equation} 
\left[\begin{array}{cc}
E_{\rm sp}+V^6_{\pi\pi}+V^{17/2}_{11}&V^{17/2}_{12}\\[1mm]
V^{17/2}_{21}&
E_{\rm sp}+V^8_{\pi\pi}+V^{17/2}_{22}
\end{array}\right],
\label{eq:mo17a}
\end{equation}
where $V^{17/2}_{21}=V^{17/2}_{12}$
is the off-diagonal matrix element
due to the $\nu\pi$ interaction and responsible for mixing the two $17/2^+$ states.
The explicit expressions in terms of the two-body $\nu\pi$ matrix elements are
\begin{align}
V^{17/2}_{11}={}&
\textstyle
\frac{3}{220}V^4_{\nu \pi}+\frac{133}{180}V^5_{\nu \pi}+\frac{95}{132}V^6_{\nu \pi}+\frac{57}{52}V^7_{\nu \pi},
\nonumber\\
V^{17/2}_{12}={}&
\textstyle
\sqrt{\frac{133}{3}}
\Bigl(\frac{3}{220}V^4_{\nu \pi}+\frac{11}{260}V^5_{\nu \pi}-\frac{5}{44}V^6_{\nu \pi}+\frac{3}{52}V^7_{\nu \pi}\Bigr),
\nonumber\\
V^{17/2}_{22}={}&
\textstyle
\frac{133}{220}V^4_{\nu \pi}+\frac{121}{260}V^5_{\nu \pi}+\frac{35}{44}V^6_{\nu \pi}+\frac{7}{52}V^7_{\nu \pi}.
\label{eq:mo17b}
\end{align}

Using the empirical $\nu\pi$ and $\pi\pi$ two-body matrix elements
and neglecting the diagonal single-particle contribution $E_{\rm sp}$,
one obtains for the energy of the $21/2^+$ level
\begin{equation}
E(21/2^+)=2.940~\text{MeV},
\label{eq:mo21c}
\end{equation}
while the matrix~(\ref{eq:mo17a}) reduces to
\begin{equation} 
\left[\begin{array}{rr}
3.040&-0.235\\-0.235&3.616
\end{array}\right]~\text{MeV},
\label{eq:mo17c}
\end{equation}
leading to the eigenvalues
\begin{equation}
E(17/2_1^+)=2.956~\text{MeV},
\quad
E(17/2_2^+)=3.700~\text{MeV}.
\label{eq:mo17d}
\end{equation}
This puts the yrast $17/2^+$ level above the $21/2^+$ isomer by 16~keV,
which is in fair agreement with the experimental energy gap of 5~keV.
Thus the main structure of these two isomers can be understood
as arising from a simple single-particle $\nu1d_{5/2}\otimes\pi0g_{9/2}^2$ configuration.
This result is in line with the  predictions of Auerbach and Talmi~\cite{auerbach1964}
and also with the conclusion of Hasegawa \textit{et al.}~\cite{hasegawa2011},
who could reproduce the order of the $17/2^+$ and $21/2^+$ levels in a LSSM calculation
and explain it qualitatively as due to the $\nu\pi$ interaction.

With the present shell-model approach
one obtains a quantitative understanding of these results.
We now understand that the dominant component of the $17/2_1^+$ state
has $J_\pi=6$, i.e. it is $|\nu1d_{5/2}\otimes\pi0g_{9/2}^2(6);17/2\rangle$,
since the mixing matrix element turns out to be quite small.
The $\nu\pi$ interaction is responsible for the inversion of the $17/2^+$ and $21/2^+$ levels.
To understand the origin of this inversion more precisely, we calculate the energy difference
\begin{align}
&E_{\rm diag}(17/2^+)-E(21/2^+)=V^6_{\pi\pi}-V^8_{\pi\pi}
\nonumber\\&\qquad
\textstyle
+\frac{3}{220}V^4_{\nu\pi}+\frac{133}{780}V^5_{\nu\pi}+
 \frac{335}{924}V^6_{\nu\pi}-\frac{199}{364}V^7_{\nu\pi},
\label{eq:mo1721}
\end{align}
where the subscript `diag' indicates that
we have taken the unperturbed (or diagonal) energy of the $17/2^+$ state
and ignored the contribution to the energy from mixing.
The difference $V^6_{\pi\pi}-V^8_{\pi\pi}$ is negative,
favoring a $17/2_1^+$ level below $21/2^+$.
However, the absolute magnitude of this difference is rather small (149~keV),
reflecting the energy spacing between the $6^+$ and $8^+$ states in $^{92}$Mo.
Because the empirical $\nu\pi$ interaction is taken from $^{92}$Nb with a $7^+$ ground state,
its contribution in Eq.~(\ref{eq:mo1721}) is {\em always} positive
since $\frac{3}{220}+\frac{133}{780}+\frac{335}{924}=\frac{199}{364}$.
The combined $\pi\pi$ and $\nu\pi$ contributions in Eq.~(\ref{eq:mo1721}) lead to an energy difference of 100~keV,
which is reduced to 16~keV by the mixing of the $17/2^+$ states.

Subtle effects in the nuclear interaction therefore play an essential role
in turning the $21/2^+$ level into an isomer
that is important in the study of isomeric depletion. On the other hand, the LSSM calculations using GLEKPN, as listed in Table II, overestimate the level energies for most of the states and do not reproduce the level inversion between ${21/2}^+$ and ${17/2}^+$ states.

Other states in $^{93}$Mo can be treated in a similar fashion.
For example, the energy of the $19/2^+$ level is
\begin{equation}
\textstyle
E(19/2^+)=E_{\rm sp}+
V^8_{\pi\pi}+\frac{7}{13}V^5_{\nu \pi}+\frac{6}{7}V^6_{\nu \pi}+\frac{55}{91}V^7_{\nu \pi},
\label{eq:mo19}
\end{equation}
with $\nu\pi$ contributions only from $J_{\nu\pi}=5$, 6, and 7.
For $J=15/2$ one finds two levels and a Hamiltonian matrix of the form 
\begin{equation} 
\left[\begin{array}{cc}
E_{\rm sp}+V^6_{\pi\pi}+V^{15/2}_{11}&V^{15/2}_{12}\\[1mm]
V^{15/2}_{21}&
E_{\rm sp}+V^8_{\pi\pi}+V^{15/2}_{22}
\end{array}\right],
\label{eq:mo15a}
\end{equation}
where
\begin{widetext}
\begin{align}
&\textstyle
V^{15/2}_{11}=
\frac{17}{550}V^3_{\nu \pi}+\frac{169}{550}V^4_{\nu \pi}+\frac{224}{325}V^5_{\nu \pi}+
\frac{228}{1925}V^6_{\nu \pi}+\frac{855}{1001}V^7_{\nu \pi},
\quad
V^{15/2}_{22}=
\frac{323}{550}V^3_{\nu \pi}+\frac{171}{550}V^4_{\nu \pi}+\frac{266}{325}V^5_{\nu \pi}+
\frac{507}{1925}V^6_{\nu \pi}+\frac{20}{1001}V^7_{\nu \pi},
\nonumber\\&\textstyle
V^{15/2}_{12}=V^{15/2}_{21}=
\sqrt{19}
\Bigl(\frac{17}{550}V^3_{\nu \pi}+\frac{39}{550}V^4_{\nu \pi}-\frac{56}{325}V^5_{\nu \pi}+
\frac{78}{1925}V^6_{\nu \pi}+\frac{30}{1001}V^7_{\nu \pi}\Bigr).
\label{eq:mo15b}
\end{align}
\end{widetext}
The off-diagonal element in this Hamiltonian matrix is very small,
leading to a negligible mixing of the two $15/2^+$ states.
With use of the empirical $\nu\pi$ and $\pi\pi$ two-body matrix elements
the two eigenvalues of the Hamiltonian matrix~(\ref{eq:mo15a}) are 3.07 and 3.50~MeV.
The first-excited $15/2^+$ state has a dominant $J_\pi=6$ component,
i.e. it is approximately $|\nu1d_{5/2}\otimes\pi0g_{9/2}^2(6);15/2\rangle$,
similar to the first-excited $17/2^+$ state
and the energy difference $E(17/2_1^+)-E(15/2_1^+)$
is therefore essentially determined by the $\nu\pi$ interaction.

Similarly, one can obtain the full spectrum
in the $\nu1d_{5/2}\otimes\pi0g_{9/2}^2$ model space.
Table~\ref{tab:enmo93} compares such calculated energies for $^{93}$Mo with the experiment
as well as with theoretical results previously reported by Hasegawa~\textit{et al.}~\cite{hasegawa2011}.
The present calculations reproduce the $5/2^+$ ground state
arising due to the odd neutron in $1d_{5/2}$
and agree fairly well with the available data
except for the $1/2^+$ and the second/third-excited $7/2^+$ levels.
The reason behind the deviations for the lower-spin levels
is the missing contributions due to excitations of the neutron
into the $2s_{1/2}$, $0g_{7/2}$, and $1d_{3/2}$ orbitals
from the $N=50$--82 valence space.
The agreement of the present calculation with the data
is even better than the LSSM calculations~\cite{hasegawa2011},
especially for the states of interest, i.e. the $17/2^+$ and $21/2^+$ isomers.
This clearly suggests the purity of the $\nu1d_{5/2}\otimes\pi0g_{9/2}^2$ configuration for these states in $^{93}$Mo. Figure~\ref{fig:enmo93} presents a comparison of our calculated energies with the experiment.
For completeness the calculated numbers drawn from Fig.~1 of Ref.~\cite{hasegawa2011} are also plotted.  

Detailed expressions for the transitions are listed in Table~\ref{tab:emmo93}
in terms of neutron and proton effective charges.
They are given in Weisskopf units (W.u.) or in units of $e^2b^4$
and with the effective charges specified in Sect.~\ref{sec:formalism}.
The $E2$ strengths for the $17/2^+\rightarrow21/2^+$ and $17/2^+\rightarrow13/2^+$ transitions
are crucial in the study of isomeric depletion.
Our calculated results are in agreement with the experimentally known value for the latter transition
while the former is estimated to be about $40\%$ reduced
in comparison to the literature value~\cite{hasegawa2011}.
The current analysis supports a nearly constant $B(E2;17/2^+\rightarrow21/2^+)$ value 
for the three choices of the neutron effective charge.
Reasonable agreement of the present $B(E2)$ predictions is found
with the data wherever available, except for the $9/2^+\rightarrow5/2^+$ transition.
This is intriguing especially since the present set of calculations
can fairly well explain the $B(E2;7/2^+\rightarrow5/2^+)$ value.
A possible explanation is that the missing contributions
of neutron excitations from below $N=50$
influence the wave function of the $9/2^+$ state.
Apparently, level energies are not much impacted by such insignificant configuration mixing
but transition probabilities are.
Future measurements of more $E2$ transition probabilities are much required. The $B(E2)$ values obtained from the LSSM calculations using the GLEKPN interaction are also consistent with the available experimental data, except for the $9/2^+\rightarrow5/2^+$ transition.

We have also obtained the $E4$ expression for $B(E4; {21/2}^+\rightarrow {13/2}^+)= (0.83 e_\nu + 3.53 e_\pi)^2$ (in $e^2b^8$) in present shell model. We utilize the measured value of $B(E4; 4^+ \rightarrow 0^+)$ in $^{92}$Mo~\cite{nndc,thesis} to fix $e_\pi \approx 1.20$ while no additional information is available for the neutron. The calculated values of $B(E4; 21/2^+ \rightarrow 13/2^+)$ are $1.25$, $1.37$, and $1.50$ W.u., obtained using neutron effective charges of $e_\nu = 1.18$, $1.48$, and $1.78$, respectively. These results compare favorably with the measured value of $1.449(17)$ W.u.~\cite{nndc}.

\begin{figure}[!htb]
\centering
\includegraphics[width=0.49\textwidth]{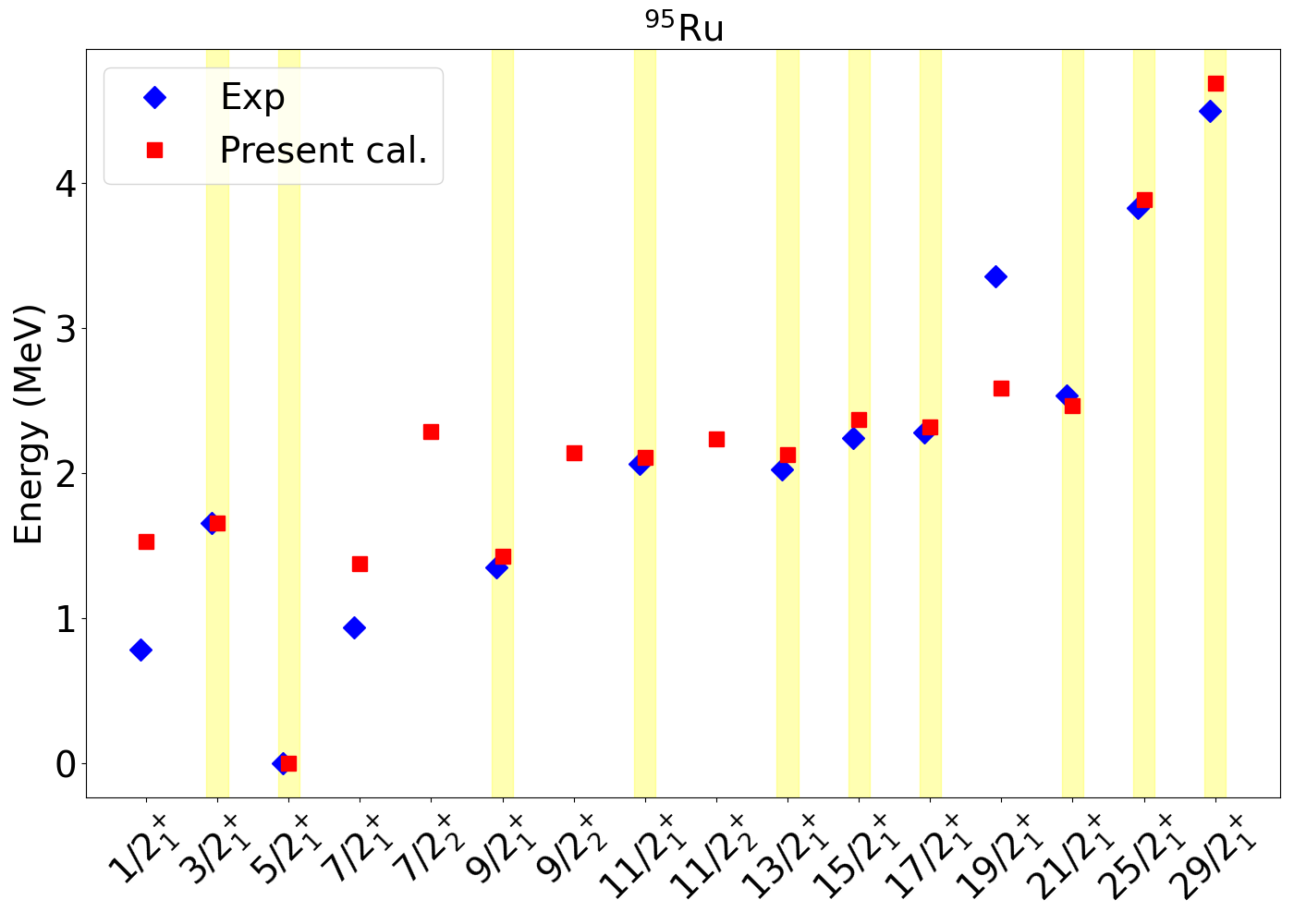}
\caption{(Color online)
\label{fig:enru95}
Experimental~\cite{nndc} spectrum of low-lying states in $^{95}$Ru
compared to the energies calculated in the $\nu1d_{5/2}\otimes\pi0g_{9/2}^4$ model space.}
\end{figure}

\begin{figure}[!htb]
\centering
\includegraphics[width=0.49\textwidth]{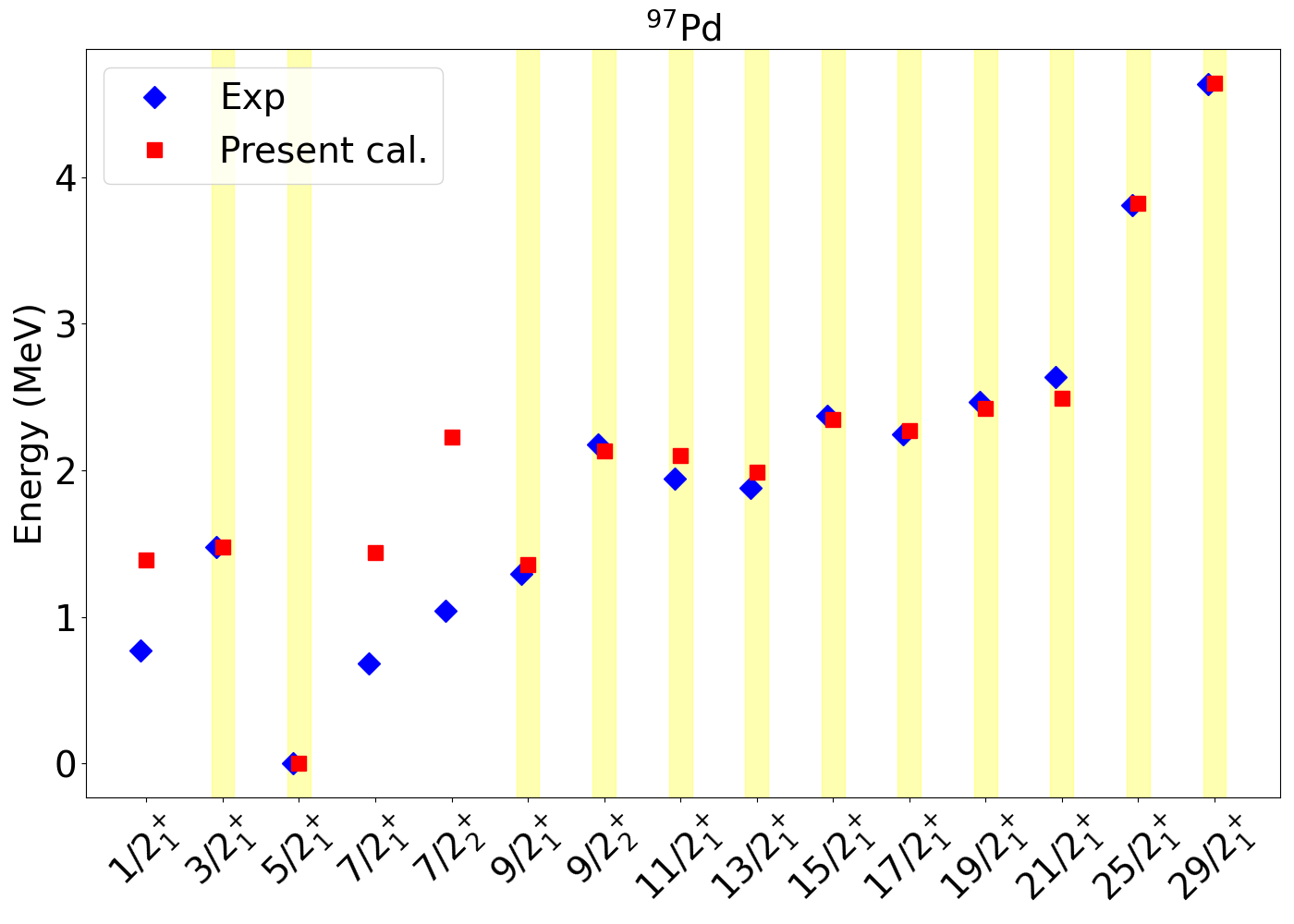}
\caption{(Color online)
\label{fig:enpd97}
Experimental~\cite{nndc} spectrum of low-lying states in $^{97}$Pd
compared to the energies calculated in the $\nu1d_{5/2}\otimes\pi0g_{9/2}^6$ model space.}
\end{figure}

\begin{figure}[!htb]
\centering
\includegraphics[width=0.49\textwidth]{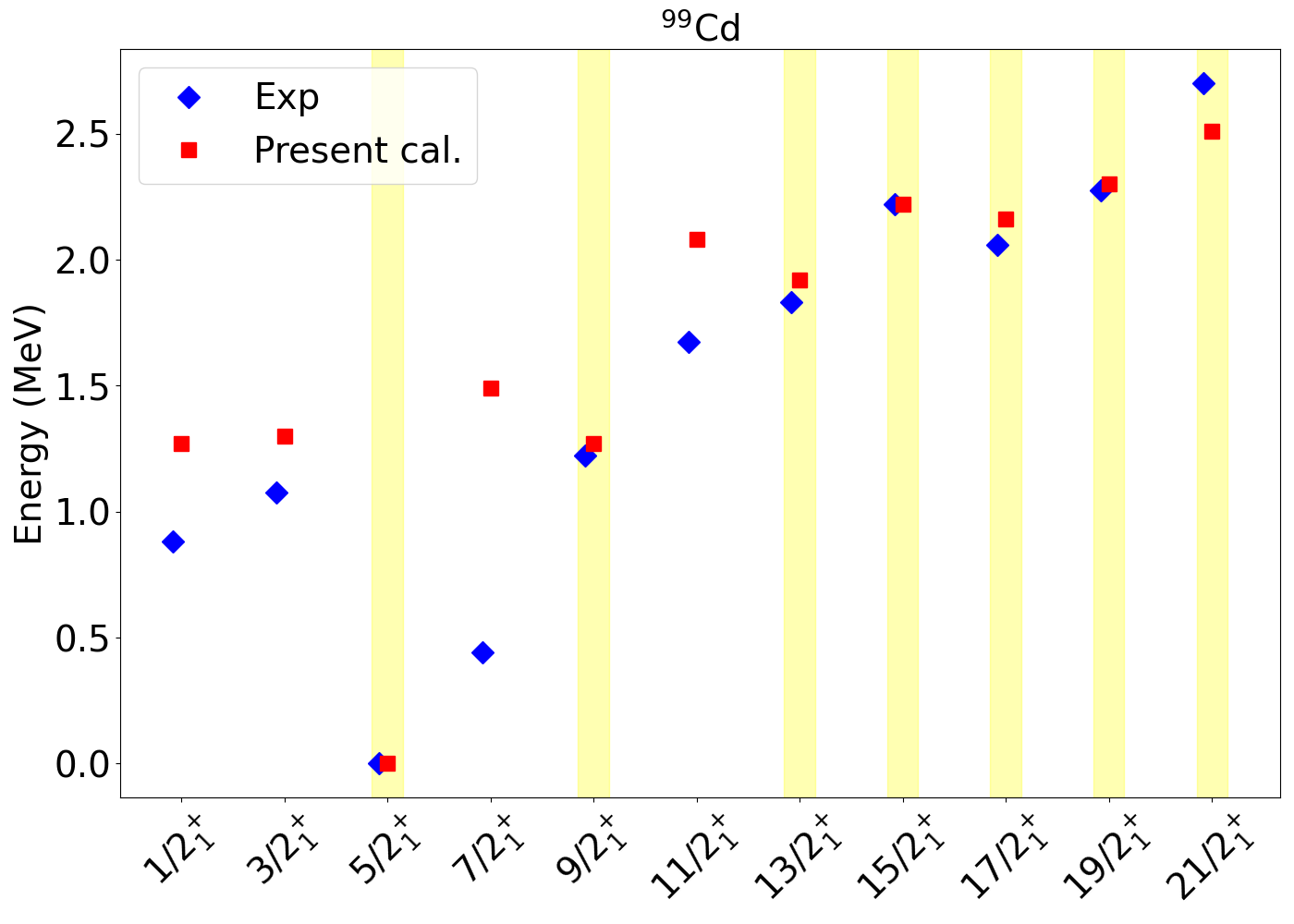}
\caption{(Color online)
\label{fig:encd99}
Experimental~\cite{nndc} spectrum of low-lying states in $^{99}$Cd
compared to the energies calculated in the $\nu1d_{5/2}\otimes\pi0g_{9/2}^8$ model space.}
\end{figure}

\subsection{The nucleus $^{95}$Ru}
\label{subs:ru95}
In the $\nu1d_{5/2}\otimes\pi0g_{9/2}^4$ model space for $^{95}$Ru
the allowed states are
$1/2^+(3)$, $3/2^+(6)$, $5/2^+(9)$, $7/2^+(10)$, $9/2^+(11)$, $11/2^+(11)$,
$13/2^+(11)$, $15/2^+ (9)$, $17/2^+ (8)$, $19/2^+ (6)$, $21/2^+ (5)$, 
$23/2^+ (3)$, $25/2^+ (2)$, $27/2^+(1)$, and $29/2^+(1)$,
where the numbers in parentheses again refer to the multiplicity of the state.
The calculated energies are listed in Table~\ref{tab:enrupdcd}
and are also plotted in Fig.~\ref{fig:enpd97} for the levels that are known experimentally.

We are particularly interested in the structure
of the $21/2^+$ and $17/2^+$ states and their relative energies.
Five states $|\nu1d_{5/2}\otimes\pi0g_{9/2}^4(\upsilon_\pi,J_\pi);21/2\rangle$
occur for $(\upsilon_\pi,J_\pi)=(2,8)$, (4,8), (4,9), (4,10), and (4,12),
where $\upsilon_\pi$ is the seniority quantum number of the four protons in $0g_{9/2}$.
The Hamiltonian matrix in this basis is
\begin{equation}
\small
\resizebox{0.30\textwidth}{!}{$\left[\begin{array}{rrrrr}
    14.04 & 0.01 & 0.09 & 0.15 & 0.08 \\
    0.01 & 15.62 & -0.19 & 0.17 & -0.12 \\
    0.09 & -0.19 & 15.90 & -0.13 & 0.05 \\
    0.15 & 0.17 & -0.13 & 15.76 & -0.13 \\
    0.08 & -0.12 & 0.05 & -0.13 & 16.74
\end{array}\right]$},
\nonumber
\end{equation} 
which shows that the lowest eigenstate
is rather pure with $(\upsilon_\pi,J_\pi)=(2,8)$
and only weakly mixes with other $21/2^+$ states.
The diagonal energy of this $(\upsilon_\pi,J_\pi)=(2,8)$ state
has contributions from the $\pi\pi$ and $\nu\pi$ interactions,
\begin{align}
&\textstyle
E_{\rm diag}(21/2^+)=E_{\rm sp}+
\frac{3}{5}V^0_{\pi\pi}+\frac{6}{11}V^2_{\pi\pi}+\frac{486}{715}V^4_{\pi\pi}
\nonumber\\&\qquad\textstyle
+\frac{87}{55}V^6_{\pi\pi}+\frac{1854}{715}V^8_{\pi\pi}+
\frac{1}{27}V^2_{\nu\pi}+\frac{49}{297}V^3_{\nu\pi}
\nonumber\\&\qquad\textstyle
+\frac{4}{11}V^4_{\nu\pi}+\frac{196}{351}V^5_{\nu\pi}+
\frac{1972}{2079}V^6_{\nu\pi}+\frac{5788}{3003}V^7_{\nu\pi},
\label{eq:ru21}
\end{align}
where the single-particle contribution to the energy is
$E_{\rm sp}=\epsilon_{1d_{5/2}}+4\epsilon_{0g_{9/2}}$.

Eight states $|\nu1d_{5/2}\otimes\pi0g_{9/2}^4(\upsilon_\pi,J_\pi);17/2\rangle$
can be constructed with $(\upsilon_\pi,J_\pi)=(2,6)$, $(4,6)^2$,  (4,7), (2,8), (4,8), (4,9), and (4,10),
where it should be noted that two seniority $\upsilon_\pi=4$ states exist for $J_\pi=6$.
The Hamiltonian matrix in this basis is
\begin{equation}
\small
\resizebox{0.475\textwidth}{!}{$\left[\begin{array}{rrrrrrrr}
   14.02 & 0.17 &   0.08 &  0.20 &  -0.12 &  -0.13 & -0.07 & -0.05 \\
    0.17 &  14.22 &  -0.00 &  -0.07 &  -0.22 &  -0.07 &  -0.02 & 0.01 \\
    0.08 &  -0.00 &  15.37 &  -0.12 &  0.03 &  -0.05 &  -0.09 & -0.08 \\
    0.20 &  -0.07 &  -0.12 &  14.96 &  -0.15 &  0.04 &  0.15 &  -0.05 \\
    -0.12&  -0.22 &  0.03&  -0.15 &  14.42 &  -0.04 &  -0.06 &  0.23 \\
    -0.13 &  -0.07&  -0.05 &  0.04 &  -0.04 &  15.53 &  0.07 & 0.06 \\
    -0.07 &  -0.02 &  -0.09 &  0.15 &  -0.06 & 0.07 &  16.09 &  0.09 \\
    -0.05 &  0.01 & -0.08 &  -0.05 &  0.23 &  0.06 &  0.09 &  15.93 
\end{array}\right]$},
\nonumber
\end{equation}
from where it can be seen that in this case the mixing is strong,
in particular between the states with $(\upsilon_\pi,J_\pi)=(2,6)$, $(4,6)$, $(4,7)$, and (2,8).
The $(\upsilon_\pi,J_\pi)=(2,6)$ state has the lowest diagonal energy given by
\begin{align}
&\textstyle
E_{\rm diag}(17/2^+)=E_{\rm sp}+
\frac{3}{5}V^0_{\pi\pi}+\frac{34}{99}V^2_{\pi\pi}+\frac{1186}{715}V^4_{\pi\pi}
\nonumber\\&\quad\textstyle
+\frac{658}{495}V^6_{\pi\pi}+\frac{1479}{715}V^8_{\pi\pi}+
\frac{233}{1512}V^2_{\nu\pi}+\frac{2347}{11880}V^3_{\nu\pi}
\nonumber\\&\quad\textstyle
+\frac{197}{616}V^4_{\nu\pi}+\frac{1709}{2808}V^5_{\nu\pi}+
\frac{12077}{10395}V^6_{\nu\pi}+\frac{4679}{3003}V^7_{\nu\pi},
\label{eq:ru17}
\end{align}
such that the difference of the lowest diagonal energies
for the $17/2^+$ and $21/2^+$ states becomes
\begin{align}
&\textstyle
E_{\rm diag}(17/2^+)-E_{\rm diag}(21/2^+)=
-\frac{20}{99}V^2_{\pi\pi}+\frac{140}{143}V^4_{\pi\pi}
\nonumber\\&\quad\textstyle
-\frac{25}{99}V^6_{\pi\pi}-\frac{75}{143}V^8_{\pi\pi}+
\frac{59}{504}V^2_{\nu\pi}+\frac{43}{1320}V^3_{\nu\pi}
\nonumber\\&\quad\textstyle
-\frac{27}{616}V^4_{\nu\pi}+\frac{47}{936}V^5_{\nu\pi}+
\frac{739}{3465}V^6_{\nu\pi}-\frac{1109}{3003}V^7_{\nu\pi}.
 \label{eq:ru1721}
\end{align}
This difference turns out to be negative,
that is, even from the diagonal energies the $17/2^+$
is predicted to lie below the $21/2^+$ level,
and the latter will not be a long-lived isomeric state.
The negative contribution of $-149$~keV to the $17/2^+$--$21/2^+$ diagonal energy difference 
stemming from the $\pi\pi$ interaction is about the same as it is in $^{93}$Mo.
In contrast, the positive contribution from the $\nu\pi$ interaction is reduced to 129~keV in $^{95}$Ru,
whereas it is 249~keV in $^{93}$Mo.
This reduction is due to the smaller coefficient of the aligned $\nu\pi$ matrix element $V^7_{\nu\pi}$ in Eq.~(\ref{eq:ru1721})
as compared to the corresponding coefficient in Eq.~(\ref{eq:mo1721}).
The strong mixing in the $17/2^+$ space further reduces the energy of $17/2_1^+$
relative to that of $21/2_1^+$ (see Table~\ref{tab:enrupdcd}),
which is in agreement with the data~\cite{nndc}.
This also underlines the important role of configuration mixing in the formation of isomers.

Table~\ref{tab:emrupdcd} lists the $B(E2)$ values that are known experimentally in $^{95}$Ru
and the corresponding results calculated in the $\nu1d_{5/2}\otimes\pi0g_{9/2}^4$ model space
or calculated with the GLEKPN interaction.
The $\nu1d_{5/2}\otimes\pi0g_{9/2}^4$ calculation predicts several $E2$ transition probabilities
that are at variance with the measured values.
The LSSM with the GLEKPN interaction does  better
and, specifically, predicts a $B(E2;21/2^+\rightarrow17/2^+)$ value that agrees with the data.
The results of both calculations agree with each other
for the $E2$ transitions between high-angular-momentum levels,
$29/2^+\rightarrow25/2^+$ and $25/2^+\rightarrow21/2^+$,
indicating that such states experience only weak configuration mixing.

For both the ${17/2}^+$ and ${29/2}^+$ states, the experimental $B(E2)'s$ considerably exceed the shell-model calculations~\cite{chowdhury1985, galindo2004} while the different shell-model calculations including the present ones are in fairly good agreement with each other. However, the lifetimes of these states present particular experimental problems, and it would be highly desirable to have new experimental data. At least for the ${17/2}^+$ state, such new data have recently been obtained~\cite{private} but details are not yet available. The states with lower angular momentum
apparently experience more configuration mixing from the orbitals
that are excluded from the simple shell-model description. In particular, the $B(E2; {11/2}_2^+ \rightarrow {7/2}^+)$ could not be explained by the present calculations.

\subsection{The nucleus $^{97}$Pd}
\label{subs:pd97}
In the $\nu1d_{5/2}\otimes\pi0g_{9/2}^n$ model space,
the nucleus $^{97}$Pd ($n=6$) is the particle-hole conjugate in the protons of $^{95}$Ru ($n=4$).
Therefore, all expressions given in Subsect.~\ref{subs:pd97} remain valid
provided we use the particle-hole $\nu\pi$ interaction of Table~\ref{tab:int}.
The results obtained in the $\nu1d_{5/2}\otimes\pi0g_{9/2}^{-4}$ configuration
are in fair agreement with the data, see Table~\ref{tab:enrupdcd},
except for the $1/2_1^+$, $7/2_1^+$, and $7/2_2^+$ levels,
as highlighted by the yellow shading in Fig.~\ref{fig:enpd97}.

The relative position of the $21/2_1^+$ and $17/2_1^+$ can be understood
with similar arguments as given in Subsect.~\ref{subs:ru95}.
The first-excited $21/2^+$ state has a dominant $(\upsilon_\pi,J_\pi)=(2,8)$ structure,
only weakly mixing with other $21/2^+$ states,
while the $17/2_1^+$ state is more strongly mixed,
the lowest diagonal energy being associated with the $(\upsilon_\pi,J_\pi)=(2,6)$ configuration.
Hence, the same difference~(\ref{eq:ru1721}) of the lowest diagonal energies
applies for the $17/2^+$ and $21/2^+$ yrast levels,
and it explains why the former lies below the latter.
Given that the $\pi\pi$ interaction is $A$ dependent (see Table~\ref{tab:int})
its contribution to the energy difference~(\ref{eq:ru1721}) is reduced to $-121$~keV.
The main change between $^{95}$Ru and $^{97}$Pd, however,
is the switch from particle-particle to particle-hole $\nu\pi$ interaction,
which reduces its positive contribution to 9~keV.

Table~\ref{tab:emrupdcd} lists the $B(E2)$ values
associated with the $17/2^+\rightarrow13/2^+$ and $21/2^+\rightarrow17/2^+$ transitions.
Only the former is known experimentally
and agrees with the simple $\nu1d_{5/2}\otimes\pi0g_{9/2}^{-4}$ estimate
but not with LSSM calculation with the GLEKPN interaction.
The predicted $B(E2;21/2^+\rightarrow17/2^+)$ value requires future measurement.   

\subsection{The nucleus $^{99}$Cd}
\label{subs:cd99}
In the $\nu1d_{5/2}\otimes\pi0g_{9/2}^n$ model space,
the nucleus $^{99}$Cd ($n=8$) is the particle-hole conjugate in the protons of $^{93}$Mo ($n=2$)
and all expressions given in Subsect.~\ref{subs:mo93} remain valid
subject to the use of the particle-hole $\nu\pi$ interaction.
The results obtained in the $\nu1d_{5/2}\otimes\pi0g_{9/2}^{-2}$ configuration
are in fair agreement with the data, see Table~\ref{tab:enrupdcd},
except for the $1/2_1^+$, $3/2_1^+$, $7/2_1^+$, and $11/2_1^+$ levels.
as highlighted by the yellow shading in Fig.~\ref{fig:encd99}.

As is the case in $^{93}$Mo, there is only a single $21/2^+$ state
while there are two allowed $17/2^+$ states $|\nu1d_{5/2}\otimes\pi0g_{9/2}^{-2}(J_\pi);17/2\rangle$
with $J_\pi=6$ and 8, respectively.
From the expressions given in Subsect.~\ref{subs:mo93},
it is found that the diagonal energies of the two $17/2^+$ states are separated by 39~keV only
and that significant mixing results from the off-diagonal matrix element of 108~keV.
The lowest $17/2^+$ eigenstate is therefore a mixture of 59\% $J_\pi=6$ and 41\% $J_\pi=8$.
Even ignoring the effect of this mixing,
from Eq.~(\ref{eq:mo1721}) one obtains a negative difference of diagonal energies,
putting the $17/2^+$ below the $21/2^+$ level.
The contribution $-147$~keV of the $\pi\pi$ interaction in $^{99}$Cd
is about the same as it is in $^{93}$Mo;
the $\nu\pi$ interaction also gives a negative contribution ($-112$~keV)
since the particle-hole matrix element $V^7_{\nu\pi^{-1}}$
is now the most repulsive one appearing in Eq.~(\ref{eq:mo1721}).

No experimental $B(E2)$ values exist in $^{99}$Cd so far (see Table~\ref{tab:emrupdcd}),
which could be a potential candidate for future measurements.

\begin{table*}[!htb]
\caption{\label{tab:emrupdcd}
Same caption as in Table~\ref{tab:emmo93}
but for $^{95}$Ru, $^{97}$Pd, and $^{99}$Cd.
Spin and parity in brackets refer to tentative assignments 
as per policy of Evaluated Nuclear Structure Data File (ENSDF)~\cite{nndc}.}
\centering
\resizebox{0.85\textwidth}{!}
{$\begin{tabular}{|c|c|c|c|c|c|c|}
\hline
&&Expression&
\multicolumn{3}{c|}{Present calculation}&GLEKPN\\
$J_{\rm i}\rightarrow J_{\rm f}$&Exp&(in units of $e^2b^4$)&
\multicolumn{3}{c|}{$e_\pi=1.32$}&$e_\pi=1.315$\\
\cline{4-6}
&&&$e_\nu=1.18$&$e_\nu=1.78$&$e_\nu=1.48$&$e_\nu=1.178$\\
\hline
$^{95}$Ru\\
\hline
$11/2_2^+\rightarrow7/2^{(+)}$ & 0.0122(15) & $(0.15\,e_\nu+1.42\,e_\pi)^2$ & 3.75  & 4.09  & 3.92 & 0.0 \\
$17/2^+\rightarrow13/2^+$ & 6.4(6) & $(0.061\,e_\nu+0.73\,e_\pi)^2$ &  0.95 & 1.02 & 0.99 & 1.9 \\
$21/2^+\rightarrow17/2^+$ & 1.94(5) & $(0.041\,e_\nu+0.41\,e_\pi)^2$ &  0.31 & 0.34 & 0.32 & 1.8\\
$(29/2)^+\rightarrow25/2^{(+)}$ & 14.5(10) & $(0.074\,e_\nu+1.41\,e_\pi)^2$ & 3.39 & 3.55 & 3.47 & 3.4 \\
$25/2^+\rightarrow21/2^+$ & $>$ 4.1 & $(0.063\,e_\nu+1.63\,e_\pi)^2$ & 4.39 & 4.54 & 4.47 & 4.8\\
\hline
$^{97}$Pd   \\
\hline
$17/2^+\rightarrow13/2^+$ & 1.5(4) & $(0.098\,e_\nu+0.75\,e_\pi)^2$ & 1.06 & 1.18 & 1.12 & 0.1\\
$21/2^+\rightarrow17/2^+$ & --- & $(0.21\,e_\nu+0.73\,e_\pi)^2$ & 1.29 & 1.57 & 1.43 & 0.5 \\
\hline
$^{99}$Cd   \\
\hline
$17/2^+\rightarrow13/2^+$ & --- & $(0.21\,e_\nu+1.15\,e_\pi)^2$ & 2.71 & 3.12 & 2.91 & 1.3\\
$21/2^+\rightarrow17/2^+$ & --- & $(0.39\,e_\nu+0.61\,e_\pi)^2$ & 1.37 & 1.93 & 1.64 & 0.3 \\
\hline
\end{tabular}$}
\end{table*}

\begin{figure}[!htb]
\centering
\includegraphics[width=0.49\textwidth]{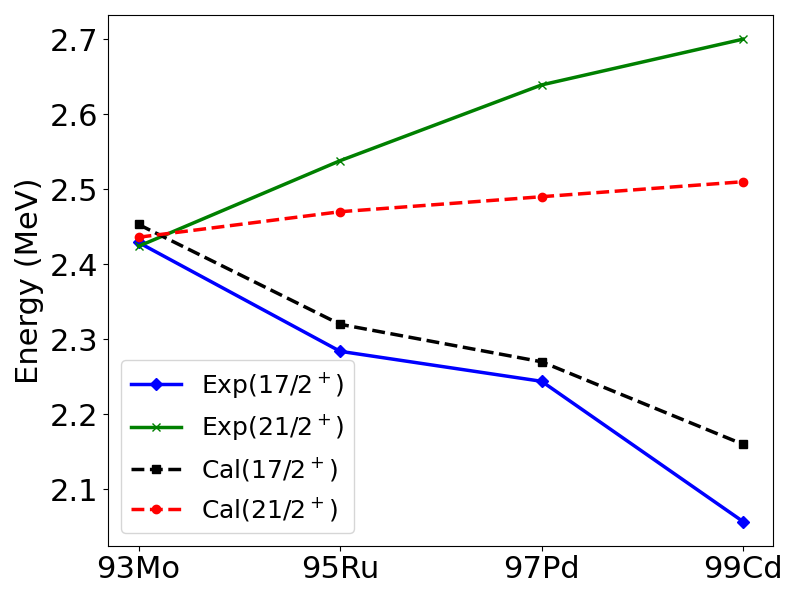}
\caption{(Color online)
\label{fig:isomer}
Experimental~\cite{nndc} and calculated energy variation of ${17/2}^+$ and ${21/2}^+$ states.
It may be noted that the $21/2^+$ level lies below the $17/2^+$ level only in $^{93}$Mo, making it a peculiar case.}
\end{figure}

\begin{figure}[!htb]
\centering
\includegraphics[width=0.49\textwidth]{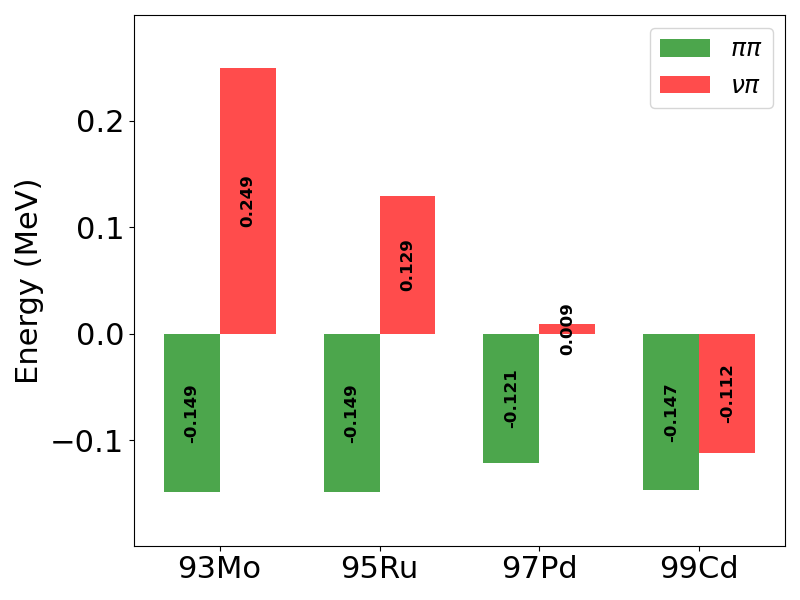}
\caption{(Color online)
\label{fig:ppnp}
The contributions of the $\pi\pi$ and $\nu\nu$ interaction to the energy difference $E_{\rm x}({21/2}^+)-E_{\rm x}({17/2}^+)$. Expressions for the diagonal energies are used, Eqs. (19) and (25), for $^{93}$Mo and $^{95}$Ru, respectively, with the $\nu\pi$ particle-particle interaction of Table I. For $^{97}$Pd and $^{99}$Cd the same equations are used but with the $\nu\pi$ particle-hole interaction.} 
\end{figure}

\section{Final Remarks}

We show in Fig.~\ref{fig:isomer} the energy variation of $17/2^+$ and $21/2^+$ levels
in the four $N=51$ isotones, compared with the theoretical results
obtained in the $\nu1d_{5/2}\otimes\pi0g_{9/2}^n$ model space.
The calculation nicely explains the experimental variation of ${17/2}^+$ with changing proton number
while it underestimates somewhat the one for $21/2^+$. The origin of this deviation is unlikely to be the $\pi\pi$ interaction since it is determined by interpolation between $^{92}$Mo and $^{98}$Cd (see Table I), which is a fairly reliable procedure. In contrast, the $\nu\pi$ interaction is obtained from $^{92}$Nb and extrapolated towards higher proton number because the spectrum of $^{100}$In is not sufficiently established. In comparing the partial information that can be extracted from the $^{100}$In spectrum with the particle-hole $\nu\pi$ matrix elements of Table I, one finds deviations up to about 100 keV, which may explain the discrepancies in excitation energies for the heavier $N=51$ isotones.

Such results provide quantitative guidance on the evolution of the $\nu\pi$ interaction shown in Fig.~\ref{fig:ppnp}, especially for the famous case of $^{93}$Mo, a NEEC candidate.
Note that the contribution of $\nu \pi$ interaction varies significantly with proton number for a fixed neutron configuration while $\pi \pi$ interaction remains nearly the same in influencing the relative energy of the diagonal components of the ${17/2}^+$ and ${21/2}^+$ states in the four $N=51$ isotones. The $\nu \pi$ interaction in $^{93}$Mo turns out to be dominating over the $\pi \pi $, resulting in the lower-lying ${21/2}^+$ state responsible for the long-lived isomeric decay, combined with the possibility of induced depopulation of the isomer by the NEEC process, as depicted in Fig.~\ref{fig:93mois}. 

\section{Conclusions}

We have studied the $N=51$ isotones $^{93}$Mo, $^{95}$Ru, $^{97}$Pd, and $^{99}$Cd, with valence protons in $0g_{9/2}$. The shell-model calculations were performed based on empirically derived two-body matrix elements to study low-energy spectra and $E2$ transition probabilities in the $\nu 1d_{5/2} \otimes \pi 0g_{9/2}^n$ configuration, $n$ being the valence proton number. The results show generally good agreement with experimental data, wherever available, and provide predictive insights for nuclei with limited spectroscopic information, which should be useful for planning future measurements.
\par

A notable outcome of our study is the demonstrated importance of the neutron-proton interaction in determining the relative energies of the ${17/2}^+$ and ${21/2}^+$ states in $^{93}$Mo. In contrast, $^{95}$Ru, $^{97}$Pd, and $^{99}$Cd exhibit reduced neutron-proton interaction strength and a shift towards configurations dominated by proton-proton correlations.
\par

Our results indicate a $\approx 40\%$ reduction in the theoretical $B(E2)$ value for the $17/2^+ \rightarrow 21/2^+$ transition in $^{93}$Mo compared to the previous estimate~\cite{hasegawa2011}. This finding is significant for interpreting nuclear excitation by electron capture (NEEC), a process relevant to the potential use of isomeric states for energy storage. It is important to note that the configuration mixing for higher-spin states such as ${21/2}^+$ to ${29/2}^+$ is governed by the chosen configuration leading to a very good agreement with the data.
\par

Level energies in $^{95}$Ru are well reproduced, though $B(E2)$ values could not be reliably determined, particularly for the ${17/2}^+ \rightarrow {13/2}^+$ transition. The presented results are also compared with large-scale shell-model calculations, denoted by GLEKPN. The problem of the $B(E2)$ estimate for the ${17/2}^+$ state in $^{95}$Ru could also not be resolved using GLEKPN and could be a subject of future measurement. For $^{97}$Pd, both level energies and $B(E2;{17/2}^+ \rightarrow {13/2}^+)$ are described successfully in our calculations, outperforming GLEKPN. Additionally, predictions are provided for the $B(E2)$ values in $^{99}$Cd.
\par

With this, we presented a quantifiable framework for identifying and assessing similar isomeric candidates in other mass regions. Future investigations could extend this methodology to other nuclei, with a particular focus on the role of higher-order multipole transitions. Notably, the ${19/2}^-$ isomeric state in $^{53}$Fe undergoes decay via an exceptionally high-multipolarity $E6$ transition. Similarly, the $M4$ transition observed for the ${19/2}^-$ isomeric state in $^{129}$Sb is of interest. In both cases, the presence of the higher-lying ${15/2}^-$ state prohibits the standard $E2$ decay path, suggesting a need for further quantifiable exploration of nucleon-nucleon interactions and understanding their impact on nuclear decay dynamics favoring the formation of peculiar isomeric candidates.





\begin{acknowledgments}
The author BM gratefully acknowledges the financial support from the HORIZON-MSCA-2023-PF-01 project, ISOON, under grant number 101150471. PW acknowledges support from the
UK Science and Technology Facilities Council under Grant
No. ST/V001108/1.
\end{acknowledgments}


\end{document}